\documentclass[%
 reprint,
 amsmath,amssymb,
 aps,
]{revtex4-2}

\usepackage{graphicx}
\usepackage{dcolumn}
\usepackage{bm}

\begin{document}

\preprint{APS/123-QED}

\title{Numerical modeling of the influence of randomly macroscopic inhomogeneities on the magnetic induction and temperature dynamics of a high-T$_c$ superconducting slab}

\author{C. Romero-Salazar}
\email{cromero.cat@uabjo.mx}
\author{P. L. Vald\'es-Negrin}
\author{C. E. \'Avila-Cris\'ostomo}
\author{O. A. Hern\'andez-Flores}%

\affiliation{%
Escuela de Sistemas Biol\'ogicos e Innovaci\'on Tecnol\'ogica \\
Universidad Aut\'onoma Benito Ju\'arez de Oaxaca\\
Av. Universidad s/n C.P. 68120 Oaxaca de Ju\'arez, Oaxaca, M\'exico
}%
\author{R. Cort\'es-Maldonado}
\affiliation{%
Tecnol\'ogico Nacional de M\'exico, IT de Apizaco,\\
Av. Instituto Tecnol\'ogico s/n, Apizaco, Tlaxcala, M\'exico C.P. 90300}
\author{V. Chabanenko}
\affiliation{Galkin Donetsk Institute for Physics and Engineering, NAS, Kyiv 03680, Ukraine}
\author{F. P\'erez-Rodr\'iguez}
\affiliation{%
Instituto de F\'isica, Benem\'erita Universidad Aut\'onoma de Puebla\\
Apartado Postal J-48 Puebla, C.P. 72570, Puebla, M\'exico
}%

\begin{abstract}
In this study we perform numerical calculations of the spatial and time variation of the magnetic induction, current density and temperature in an infinite superconducting slab with macroscopic inhomogeneities. We also modeled an homogeneuos superconductor as a referential material, considering characteristic magnetothermal properties of a high-T$_c$ single crystal. The theoretical formalism is enclosed in the dynamical regime and a continuum electrodynamic approach. In particular, we studied the correlation between the magnetic jumps and temperature rise occurrence with the presence of macroscopical inhomogeneities, modeled through an stochastic current-carrying ability spatial profile.

\end{abstract}

\maketitle

%
\section{Introduction}
The goals to investigate the vortex dynamics of type-II superconductors are to understand the physical mechanisms of their magnetization, to improve its critical current density performance, to predict the flux penetration dynamics and the conditions where, an abrupt increment of the magnetic flux, temperature or mechanical stress, can originate undesirable consequences as thermal quenching or permanent damage of the sample \cite{Gurevich2001,Wenger2001,Ainslie_2015}. 

For practical applications, bulk superconductors with a high-T$_c$ or a MgB$_2$ single-grain as seed, can be employed as trapped field magnets or passive magnetic shields \cite{Hull_2003,Durrell_2018, Gozzelino_2019}. Metallic sample holders, sheaths, or tubes placed around the samples are used experimental and theoretically as passive quench protectors or to reduce tensile stress \cite{Ainslie_2019,Ainslie_2016,ChyuCryo1992,YangJAP2012}.
Due to the nature of its growth conditions, these materials have inhomogeneities as the growth sector boundaries and, like superconductors obtained with different techniques, they have irregular borders, micro- and macro-cracks, subgrain boundaries, dislocations, oxygen-deficient regions and non-superconducting particles \cite{Eisterer_2003,Durrell_2018,Lojka2020}. 

The numerical modeling of magnetic, thermal and mechanical properties of high-T$_c$ superconducting materials has become a necessary and viable way to interpret experimental results and to predict their current-carrying ability performance \cite{Ainslie_2015,Xia_2017,Youhe2020}. In particular, the electromagnetic and thermal properties of superconductors with macroscopic defects have been numerically emulated through an inhomogeneous critical current density  \cite{Gurevich1999,Ainslie_2014,RomeroAPA2016,RomeroJAP2017,Hirano_2020}.

In this paper we focus on the modeling of the influence of macroscopic inhomogeneities on the temperature distribution in a high-T$_c$ superconductor subjected to a time-varying external magnetic field, considering the conventional bath cooling with liquid nitrogen and the lower cryocooling with helium exchange gas. For this purpose, we solved numerically the Maxwell equations  coupled with the heat diffusion equation at the dynamic regime. We considered that the temperature varied in one direction only, however, its spatial variation is not negligible and it varies non-uniformly with time.  We used as reference the case of an homogeneous sample with characteristic magnetothermal properties of a high-T$_c$ superconductor, specifically, we used experimental information of large YBa$_2$Cu$_3$O$_7$ single crystals \cite{Guillot1988,FUJIYOSHI19951609}.

\section{Theoretical framework} 
Consider an infinite inhomogeneous type-II superconducting slab of thickness $d$, with dimensions
 $|y|<\infty, |z|<\infty$ and $0 \leq x\leq d$. The sample interacts with an external homogeneous-time-varying external magnetic field, parallel to the $z-$axis, with magnitude $H_a=(rr)t$ where $rr$ is its ramp rate. In this so-called parallel geometry, the magnetodynamics is described by Faraday's law:

\begin{equation}\label{TF_ec1}
\frac{\partial Bz}{\partial t} = \frac{\partial}{\partial x}
\left(\frac{\rho}{\mu_0} \frac{\partial B_z}{\partial x} \right),
\end{equation}

with the initial (IC) and boundary (BC) conditions  
\begin{align*}
\text{IC:}  	& &B_z(x,t=0) =0, 	\\ 	
\text{BC:} 	& &B_z(x=0,t) = B_z(x=d,t) = \mu_0H_a.				
\end{align*}

The equation (\ref{TF_ec1}) is obtained combining the constitutive relation
$\mathbf{E} = \rho \mathbf{j}$ and Amp\`ere's law $\mu_0\mathbf{j} =- \partial B_z/\partial x\hat{\mathbf{y}}$, here it is assumed the resistivity $\rho = \rho(B_z,T)$ as a scalar function depending on the magnetic induction and temperature. We use the standard linear relationship $B(H)=\mu_0 H$ because we assume that the lower critical field satisfies the condition $H_{c1}\ll H_a$, therefore, geometrical barrier effects and the Meissner currents can be neglected, additionally, due to the infinite geometry, it is not necessary to contemplate the demagnetization field.

To monitor the spatial and time variation of the temperature, we solve the one-dimensional heat diffusion equation

\begin{equation}\label{TF_ec2}
\rho_mC\frac{\partial T}{\partial t} = \frac{\partial}{\partial x}
\left(\kappa\frac{\partial T}{\partial x} \right) + \frac{\partial Q}{\partial t}.
\end{equation}
$\rho_m$ is the mass density, $C$ is the specific heat, and $\kappa$ is the thermal conductivity; the heat generated in the penetrated region of the superconductor is given by Joule's law $\partial Q/\partial t=\mathbf{E} \cdot \mathbf{j} =\rho j_y^2$. 
Here we model a practical situation where the transient heat-conduction problem is linked to convective boundary conditions at the surface of the superconductor \cite{Holman}. We perform numerical experiments taking into account that the sample is in a cryogenic thermal bath, considering heat transfer from the surface of the superconducting sample to the surrounding fluid by convection, then the balance of the thermal energy transfer rate leads to the conditions:
\begin{align*}
\text{IC:}  	& &T(x,t=0) =T_B, 	\\
\text{BC:} 	& &\mathbf{\hat{n}}\cdot (-\kappa\nabla T)=h^{*}(T^{\gamma} - T_B^{\gamma}).  				
\end{align*}

The initial condition (IC) corresponds to an experiment starting after the sample is cooled down up to the desired temperature $T_B<T_c$, with the external magnetic field turned off. Eventhough the complexity of the physical mechanism of convection, $h^{*}$ can be a fixed empirical constant or a function of the temperature. We consider a constant $h^*$ and an ideal contact between the slab surface and the cryogenic fluid. 

The equations (\ref{TF_ec1}) and (\ref{TF_ec2}) (with their corresponding IC and BC), together with proper phenomenological functions for $\rho$, $\kappa$, $C$ and the critical current density $j_c$ describe the magnetothermal dynamics in the superconducting material.
\section{Dimensionless equations} 
To perform numerical calculations is convenient to work with dimensionless equations. The main physical quantities are the position $x$, time $t$, temperature $T$, and magnetic induction $B_z$. Each one can be defined in terms of characteristic values, as follows:
\begin{align*}
 x = (d/2)\bar{x}, &	&  t = t_0\bar{t},&	&T	= T_c\bar{T},	& & B_z = B_0\bar{B}_z. 	 		
\end{align*}
Here we use as characteristic values: the sample thickness $d$, $t_0=1$s, the critical temperature of the superconductor $T_c$, and $B_0=\mu_0d\, j_0$, $j_0$ is the maximum value of the critical current density that can mantain an homogeneous superconductor without dissipation, extrapolated to $T=0$.

The phenomenological models are expressed as dimensionless functions:
\begin{align*}
\text{Resistivity}  					&  	& \rho = \rho_0\bar{\rho}(\bar{B_z},\bar{T}) 			\\ 	
\text{Thermal conductivity} 	&	& \kappa = \kappa_0\bar{\kappa}(\bar{T})	\\ 	\text{Volumetric heat capacity} & & c = \rho_mC = c_0\bar{c}(\bar{T}) 	
\end{align*}

The equations (\ref{TF_ec1}) and (\ref{TF_ec2}) without dimensions, with their respective initial and boudary conditions, are
\begin{eqnarray}\label{TF_ec1a}
\frac{\partial \bar{B}_z}{\partial \bar{t}} 
& = & \frac{t_0}{t_1}\frac{\partial}{\partial \bar{x}} 
\left(\bar{\rho}\frac{\partial \bar{B}_z}{\partial \bar{x}} \right)  
\end{eqnarray}
\begin{align*}
\text{IC:}	&&  \bar{B}_z(\bar{x},\bar{t}=0)=0,   		\\  
\text{BC:}	&&   \bar{B}_z(\bar{x}=0,\bar{t}) = \bar{B}_z(\bar{x}=2,\bar{t}) = \bar{B}_a,  	
\end{align*}
\begin{eqnarray}\label{TF_ec2a}
\bar{c}(\bar{T})\frac{\partial \bar{T}}{\partial \bar{t}} 
& = & \frac{t_0}{t_2}\frac{\partial}{\partial \bar{x}}
\left(\bar{\kappa}(\bar{T})\frac{\partial \bar{T}}{\partial \bar{x}} \right) + 
\frac{t_0}{t_3}\bar{\rho}(\bar{B}_z,\bar{T})\left( \frac{\partial \bar{B}_z}{\partial\bar{x}} \right)^2 
\end{eqnarray}
\begin{align*}
\text{IC:}  	& &\bar{T}(\bar{x},\bar{t}=0) =\bar{T}_B, 	\\
\text{BC:} 	& &\mathbf{\hat{n}}\cdot (-\bar{\kappa}\nabla\bar{T})=h(\bar{T}^{\gamma} - \bar{T}_B^{\gamma}),  				
\end{align*}

here $h=h^{*}\,d\,T_c^{\gamma-1}/ \kappa_0$. There are three characteristic times $t_1$, $t_2$ and $t_3$ defined as 
\begin{align*}
t_1 = \frac{\mu_0d^2}{\rho_n}, &	  &t_2 = \frac{c_0d^2}{\kappa_0}, &	&
t_3 = \frac{\mu_0^2d^2c_0T_c}{\rho_0 B_0^2} = t_1\frac{c_0T_c}{B_0H_0}.	
\end{align*}
 $t_1$ is the time in which a magnetic profile is formed at a characteristic distance, $t_2$ is the characteristic time of the thermal response, that is, how long the heat flux takes to transport from the boundaries of the superconductor to the center of the sample, $t_3$ defines the moment when the magnetic energy exceeds the thermal energy, causing an increment of the temperature.
\section{Phenomenological equations} 
In this section we present the dimensionless version of the phenomenological equations required to
model the superconducting properties of the sample under study, whose main feature is that it has macroscopic inhomogeneities.

First, we use a highly non-linear $E=\rho j$ relationship for type-II superconductors, where the resisitivity $\rho=\rho_0\bar{\rho}(\bar{B}_z,\bar{T})$ is given by the equation

\begin{equation*}
	\rho= \left\{
\begin{array}{c l}
  \rho_{pl}\rho_n/(\rho_{pl}+\rho_n), & T <T_c  \\
  \rho_n, & T \geq T_c.
\end{array}
\right.
\end{equation*}
The resistivity is modeled as a power-law with
\begin{equation*}
\rho_{pl}=\frac{E_c}{j_c}\left(\frac{j}{j_c}\right)^{n-1},
\end{equation*}
where $n$ is the flux creep exponent, $E_c$ is obtained with the voltage criterion by transport current measurements, and $j_c=j_c(B,T)$ is the critical current density depending on the magnetic induction and  temperature. We chose the resistivity of the superconductor at normal state as characteristic value $\rho_0=\rho_{n}$ \cite{Youhe2020}. 

Then, the equation for the dimensionless resistivity is

\begin{equation}\label{Rhoa}
	\bar{\rho}= \left\{
\begin{array}{c l}
  \bar{\rho_{pl}}/(1+\bar{\rho_{pl}}), & \bar{T} <1 \\
  1, & \bar{T} \geq 1,
\end{array}
\right.
\end{equation}

The critical current density is of fundamental importance to describe the electrodynamic properties of a superconducting material because it determines the threshold between dynamic and stationary states. Here it is modeled as $\bar{j}_c=\bar{j}_{c0}\, \bar{j}_T(\bar{T})\bar{j}_B(\bar{B})$ \cite{KHENE1999727, PhysRevB.74.054507}, where $\bar{j}_c=j_c/j_0$, $\bar{j}_{c0}=j_{c0}/j_0$, and

\begin{equation}\label{jcTBa}
\bar{j}_T (\bar{T}) =  \left(1-\bar{T}^2 \right)^{3/2}, \quad
\bar{j}_B (\bar{B})  =  \frac{1}{\left(1+\bar{B}/\bar{B}^* \right)^m}.	
\end{equation}
$\bar{B}^*$ and the power $m$ are fitting parameters. 
Finally, we use the dimensionless thermal conductivity and heat capacity functions:
\begin{equation*}
\bar{\kappa}  =  \bar{T}^3, \quad	
\bar{c} 		  =  \bar{T}^3. 		
\end{equation*}
These functions are valid to describe the thermal properties of high-T$_c $ superconductors at low temperatures (below the nitrogen liquid temperature). However, to perform numerical simulations of the magnetothermal instabilities, the literature suggests to know such functions in a broad temperature range, for this purpose, an interpolation of experimental curves is required \cite{FUJIYOSHI19951609}.

\subsection{Model for the critical current density of a sample with macroscopic inhomogeneities}
\begin{figure}[htbp!]
\includegraphics[width=1\linewidth,height=\textheight,keepaspectratio]{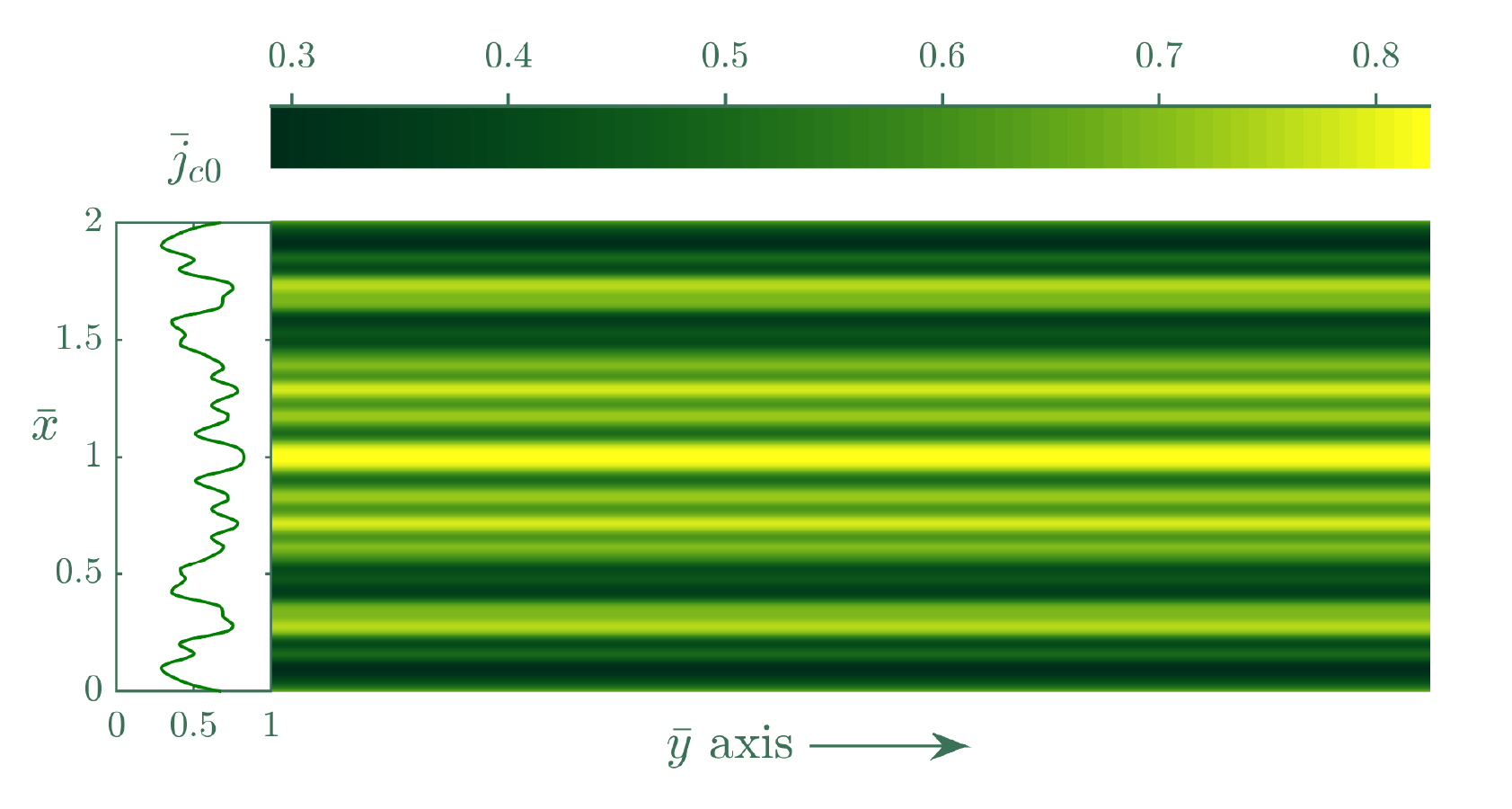}
\caption{Pseudocolor plot of the randomly distributed $\bar{j}_{c0}(\bar{x})$ magnitude, across the thickness of an infinite superconductor ($|\bar{y}|<\infty, |\bar{z}|<\infty$ and $0 \leq \bar{x}\leq 1$). The stochastic profile is symmetric with respect of the center of the sample. \label{Efig1}}
\end{figure}

Critical current density models require a set of parameters which are fitted according to the material under study, if the superconductor is anisotropic with respect of its current-carrying ability, like the high-T$_c$ superconductors, one can estimate the $j_{0}$ value from transport current or magnetization measurements along different directions of the sample. In our study we assume that the external magnetic field is parallel to the principal $c-$axis of the crystal, thus the superconducting currents flow in the isotropic $a-b$ plane, in one direction. 
Additionally, since our description is at the macroscopic level, both the electromagnetic properties and the inhomogeneities are averaged quantities. The macroscopic inhomogeneities are modeled as a randomly spatial profile, along the $x-$axis, of the maximum current-carrying ability of the sample governed by
\begin{equation}
\bar{j}_{c0}(\bar{x}) = 1-F(\bar{x}),
\end{equation}
where $0<F(\bar{x})<1$ is an stochastic profile, given the geometry of the sample, it should be symmetric with respect to the center of the sample. Figure \ref{Efig1} shows the pseudocolor plot of a randomly distribution of the $\bar{j}_{c0}(\bar{x})$ magnitude.
For the numerical simulations, the stochastic profile is constructed as follows:
\begin{enumerate}
 \item{A set of $N$ elements ${\bar{x}_1,\bar{x}_2,\dots ,\bar{x}_N}$ is defined.}
 \item{A random number $\xi_k=F(\bar{x}_k)$, distributed uniformly between zero and one, is chosen.} 
\item{A piecewise polynomial structure is constructed by means of a pchip MATLAB function (pchip means a piece cubic Hermite interpolating polynomial), such structure is evaluated at the query points.}
\item{The random profile is symmetric respect to the center of the sample.}
\end{enumerate}
\section{Results} 
The numerical experiments started emulating zero field cooling conditions $\bar{B}_z(\bar{x},\bar{t}=0)=0$ with the sample surrounded by a cryogenic fluid $\bar{T}(\bar{x},\bar{t}=0)=\bar{T}_B$. All the experiments used a ramp rate ({\it rr}) of $3\times10^{-2}$ T/s to increase the external field from zero up to a maximum value $\bar{B}_{max}=7$. For the thermal bath we selected two usual working temperatures  $T_B=4.2$K, $77$K ($\bar{T}_B=0.0467, 0.8556$, respectively). To model the inhomogeneous sample, only one $\bar{j}_{c0}(\bar{x})$ profile was employed. All the numerical calculations were performed considering a plate of thickness $d=2.65 \times 10^{-3}$m, and the Kim-Anderson parameters $\bar{B}^*=0.75$ and $m=1$. 

To analyze the effect of the macroscopic inhomogeneities on the magnetothermal dynamics of a high-$T_c$ superconducting sample, we used experimental information of large YBa$_2$Cu$_3$O$_7$ single crystals. Characteristic magnetic and thermal values of a YBa$_2$Cu$_3$O$_7$ single crystal or (RE)BCO bulk superconductors (where RE$=$rare earth or Y) are: $T_c=90$K, $\rho_n=1 \times 10^{-6}\Omega \cdot$m, $\kappa_0= 20$ W/(m$\cdot$K), $c_0=1\times10^{6}$ J/(Kg$\cdot$K), $E_c=1\times10^{-4}$ V/m  \cite{Duron2004,Ainslie_2014}. We chose $t_0=1$s, $B_0=1$T, $j_0=50B_0/(\mu_0d)$, and the creep exponent $n=90$. The parameters $h=0.005$ and $\gamma=1$ correspond to a Newton's law for convection.

We employed the method of lines to solve numerically the set (\ref{TF_ec1a})-(\ref{jcTBa}), such methodology transforms the partial differential equation into an initial-value ordinary-differential equation system by a spatial discretization of (\ref{TF_ec1a}), using a second-order accurate spatial discretization based on a fixed set of nodes, in this work we used an xmesh of 120 nodes. The ordinary differential equation system is a stiff system \cite{Shampine:2007}, so the integration is performed with implicit methods due to its excellent numerical stability. We wrote the codes and ran our program in the software MATLAB, the solver {\it ode15s} --a variable order solver based on numerical differentiation formulas--was required. 

The magnetic induction $\bar{B}_z(\bar{x})$ profiles are specular with respect to the center of the sample (located at $\bar{x}=1$), and the current density $j_y$ profiles are antisymmetric with respect to the center of the sample, then, we present only half of each profile because they contain all the physical information required to analyze their dynamical behavior as the external field $\mathbf{H}_a$ is increased.

To study the effect of the macroscopic inhomogeneities we compared the magnetic induction, current density and temperature distributions of an homogeneuos sample with the corresponding distributions of the inhomogeneous one. 

\begin{figure}[htbp!]
\includegraphics[width=1\linewidth,height=\textheight,keepaspectratio]{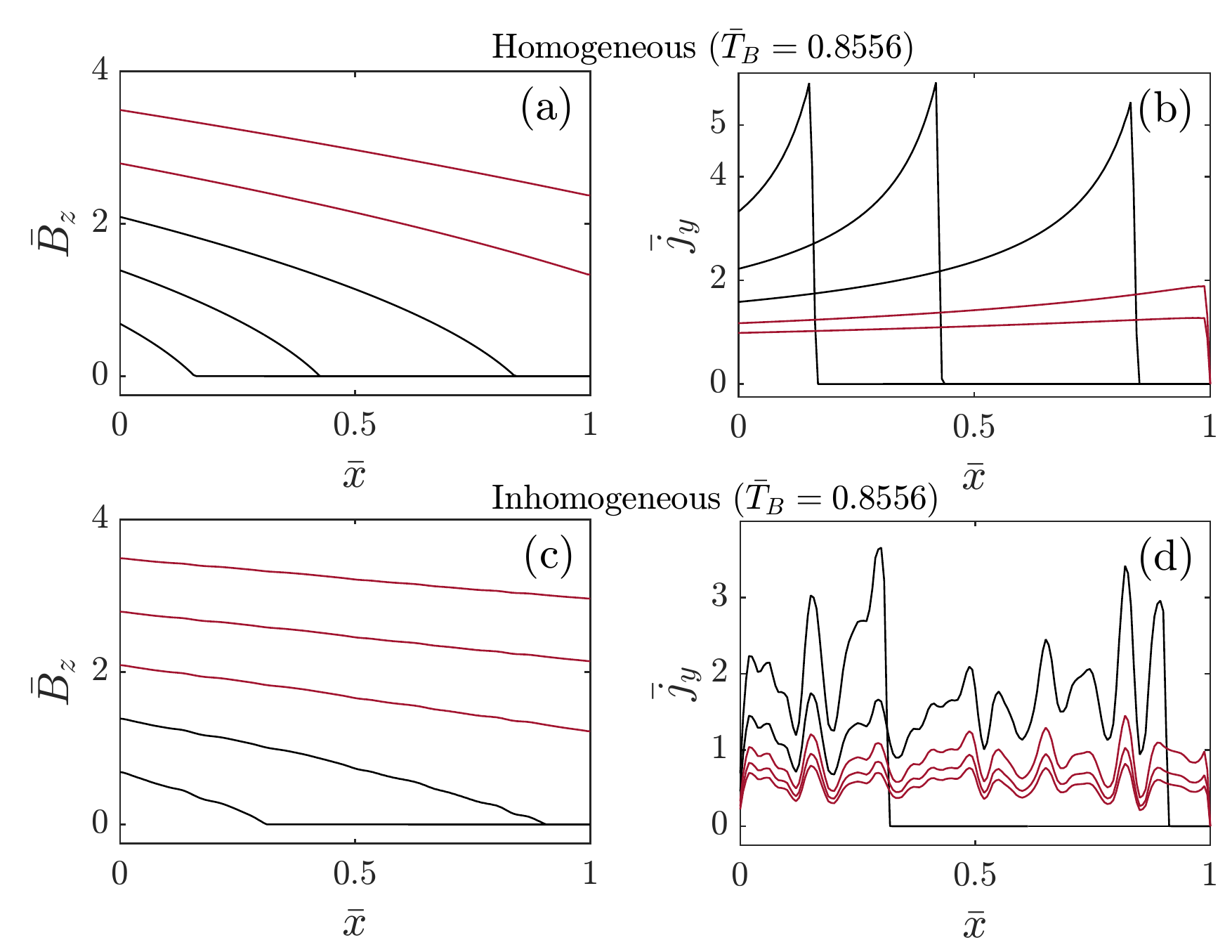}
\caption{Set of $\bar{B}_z(\bar{x})$ and $\bar{j}_y(\bar{x})$ profiles calculated numerically, as the external field was increased, of the homogenous and inhomogeneous samples in a liquid nitrogen bath ($T_B=77K$ or $\bar{T}_B=0.8556$). The black profiles correspond to partial penetrated states, the red ones belong to full penetrated states.\label{Rfig1}}
\end{figure}

Figure \ref{Rfig1} shows a set of $\bar{B}_z(\bar{x})$ and $\bar{j}_y(\bar{x})$ profiles of the samples in the liquid nitrogen bath ($T_B=77K$ or $\bar{T}_B=0.8556$). The black profiles correspond to partial penetrated states and the red ones belong to full penetrated states. Panels (a) and (b) correspond to the homogeneous superconductor, here, the profiles evolution is characteristic of the high temperature case without flux jumps, as the experimental evidence for single-crystal and bulk YBa$_2$Cu$_3$O$_7$ samples. In panels (c) and (d) is evident the influence of the inhomogeneities on the material response. For the same magnitudes and ramp rate of the external magnetic field used to study the homogeneous sample, it was obtained that the inhomogeneities promoted a full penetration into the sample at smaller values than in the homogeneous case, compare panels (a) and (c) of Fig.~\ref{Rfig1}. 

The $B_z$ profiles of the inhomogeneous superconductor presented ripples due to the randomness of the current-carrying ability $\bar{j}_{c0}(\bar{x})$, since the current density is connected with the induction field through Ampere's law, the $\bar{j}_y(\bar{x})$ profiles show a complex nonmonotonic behavior with magnitude smaller than the homogeneous sample, see panels (b) and (d) of Fig.~\ref{Rfig1}.

After the final state (the upper red $\bar{B}_z$ profiles in panels (a) and (c)) if the external field is turned off, we expect that the inhomogeneous sample will trap more field than the homogeneous one because the magnetic induction had a more uniform distribution through the sample. One can especulate then that the inhomogeneous sample could be used it as a trapped field magnet, besides, it did not present magnetothermal instabilities at this thermal bath.

\begin{figure}[htbp!]
\includegraphics[width=1\linewidth,height=\textheight,keepaspectratio]{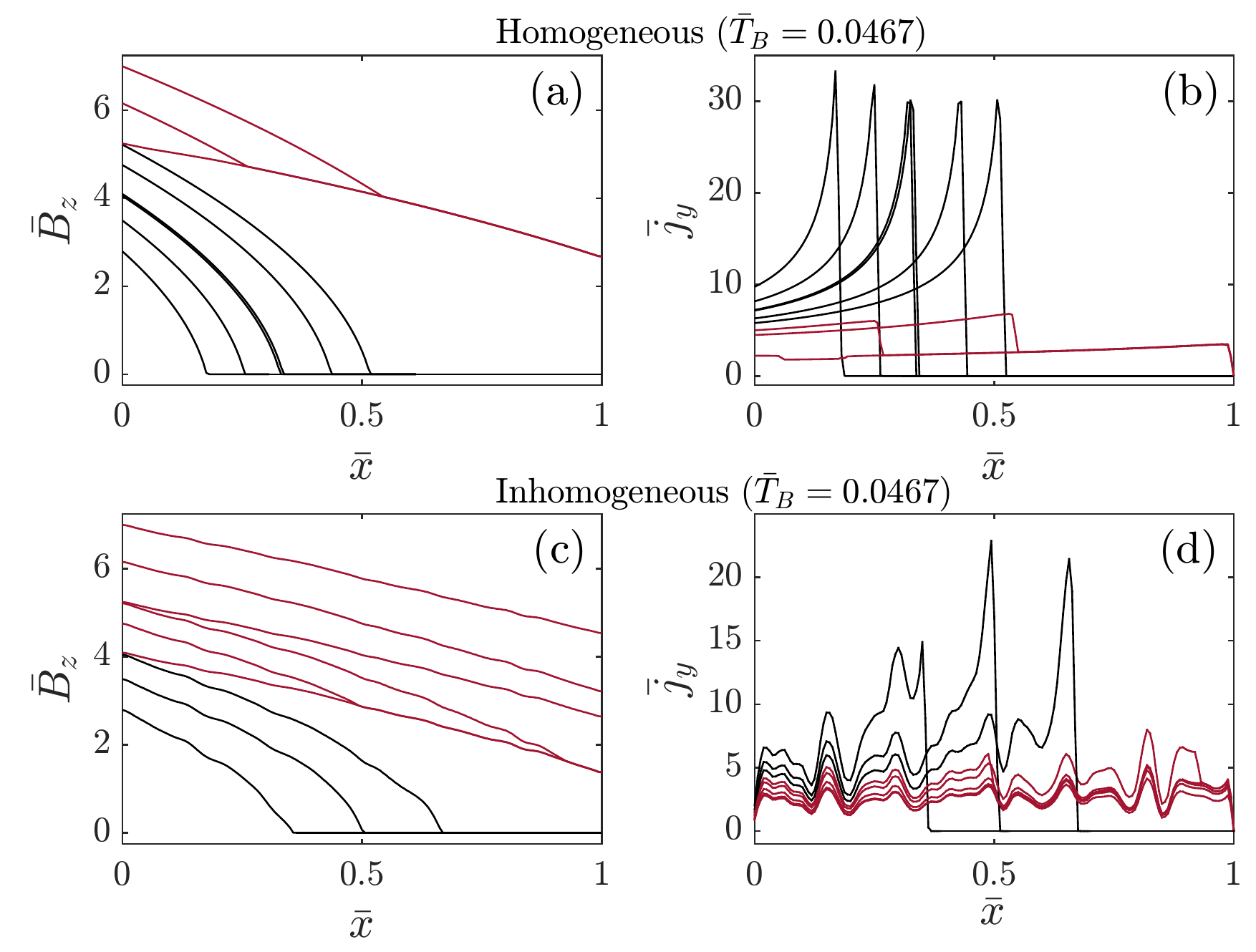}
\caption{Set of $\bar{B}_z(\bar{x})$ and $\bar{j}_y(\bar{x})$ profiles calculated numerically, as the external field was increased, when the samples were in a helium gas environment ($T_B=4.2K$ or $\bar{T}_B=0.0467$). The black $\bar{B}_z(\bar{x})$ and $\bar{j}_y(\bar{x})$ profiles correspond to states before a flux jump and the red ones to states during or after a flux jump. \label{Rfig2}}
\end{figure}

We modeled also the case of the samples in a helium gas environment ($T_B=4.2K$ or $\bar{T}_B=0.0467$), at this temperature a superconducting material used to be unstable, thermomagnetically speaking. The black $\bar{B}_z(\bar{x})$ and $\bar{j}_y(\bar{x})$ profiles correspond to states before a flux jump occurs, and the red ones to states during or after a flux jump.  As the external field was increased, one flux jump occurred in the homogeneous sample, as can be seen at panel (a) of Fig.~\ref{Rfig2}.  
At this low temperature, due to the magnetic and thermal characteristics of the samples (modeled through the critical current density $\bar{j}_c(\bar{B},\bar{T})$), the penetration of the external field required higher
intensities comparing with the case of the sample in a thermal bath of $T_B=77$K. The currents are one order of magnitude larger than the case of the sample at $T_B=77$K, but decay abruptly due to the flux jump.

In the inhomogeneous material occurred two flux jumps, see panel (c) of Fig.~\ref{Rfig2}, in both samples, as it was expected, each flux jump promoted an abrupt flux penetration. Since the inhomogeneities act as a pinning landscape, the external field penetrates into the sample at lower values, comparing with the homogeneous one. 
The rippled-shape of the magnetic induction, see panel (c), is more pronounced due to the lower temperature, this effect is diminished as the external field becomes higher. On the other hand, a higher temperature of the thermal bath ($T_B=77K$) tends to smooth the magnetic induction profiles.  
The inhomogeneities reduce the current density comparing with the homogeneous case, before and after the flux jump, as can be noticed comparing panels (b) and (d). As the experimental evidence, the flux jumps promote the decrease of  $\bar{j}_y(\bar{x})$ for both samples, however, this effect is magnified in the inhomogeneuos sample. This is the reason why the slope of the magnetic induction profiles be also smaller if they are compared with the corresponding profiles of the homogeneous case. 
Since we are also interested to know what is the role of the inhomogeneities and how the magnitude of the external magnetic field affects the temperature dynamics in the superconductor, it was calculated the spatial average of the temperature as the external field was increased. 

\begin{figure}[htbp!]
\includegraphics[width=1\linewidth,height=\textheight,keepaspectratio]{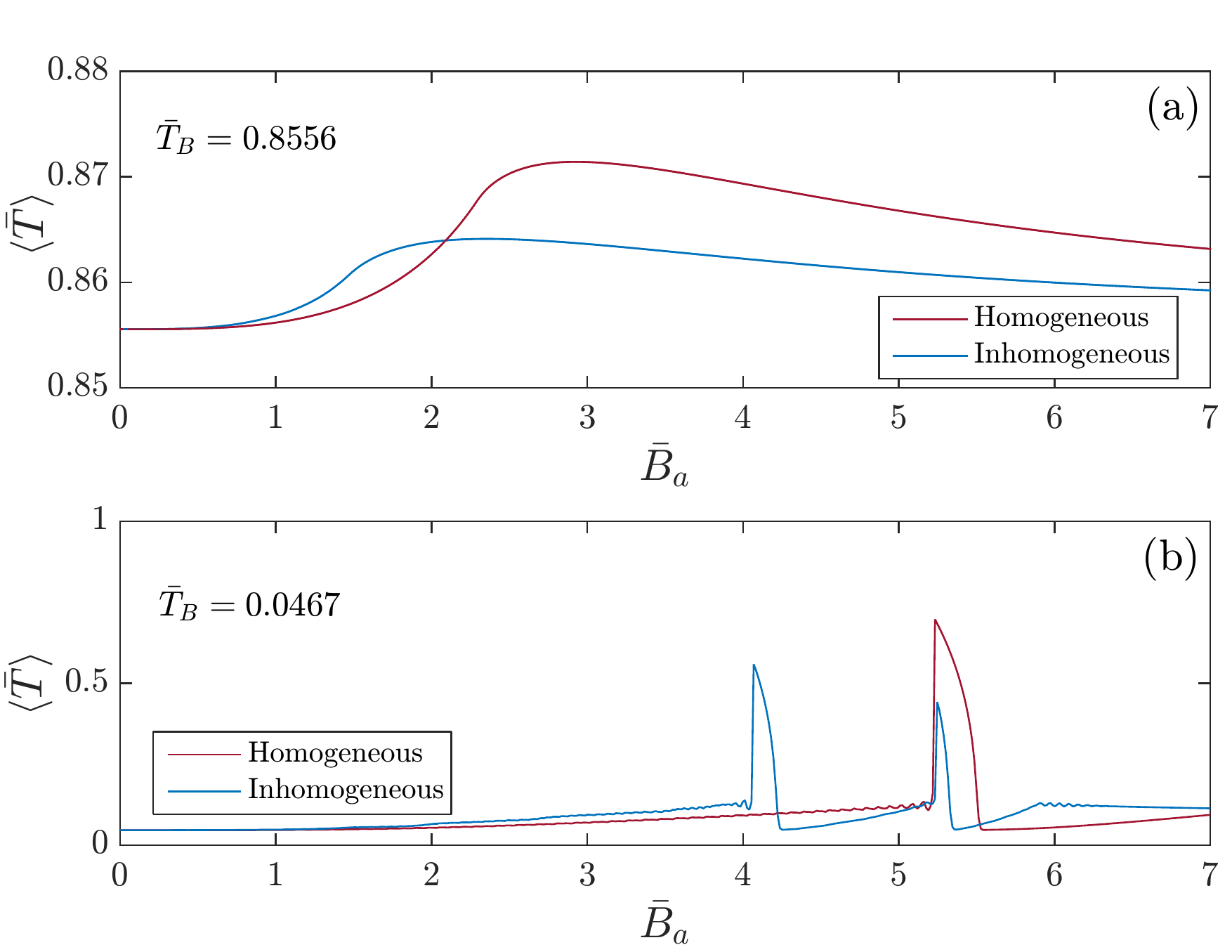}
\caption{Spatial average of the temperature $\langle\bar{T}\rangle$ as the external field was increased. Panel (a) corresponds to the case when the two samples were in a nitrogen thermal bath. The two curves presented at panel (b) were calculated at the same conditions of panel (a), except that  the temperature bath was $\bar{T}_B=0.0467$. \label{Rfig3}}
\end{figure}

Fig.~\ref{Rfig3} shows the spatial average of the temperature $\langle\bar{T}\rangle$ as the external field increased from zero up $\bar{B}_a=7$.
Panel (a) corresponds to the case when the samples were in a nitrogen thermal bath, according to the experimental evidence, in both simulations the increment of the temperature was not relevant. The two curves presented at panel (b) were calculated at the same conditions of panel (a), except that the temperature bath was $\bar{T}_B=0.0467$. The inhomogeneuos sample presented two sharp increments of the temperature, meanwhile the homogeneous presented only one of higher size. To explain these results is beyond the scope  of this study.

The temperature dynamics obtained numerically was in accordance with previous experimental and theoretical works with superconducting films and melt textured YBa$_2$Cu$_3$O$_7$ samples where the magnetothermal instability was manifested as a quasiperiodic oscillatory evolution of the temperature  \cite{Legrand1993,Yang2010,Vestgarden2016}, however, in our calculations an stability criterion was not required.

\begin{figure}[htbp!]
\includegraphics[width=1\linewidth,height=\textheight,keepaspectratio]{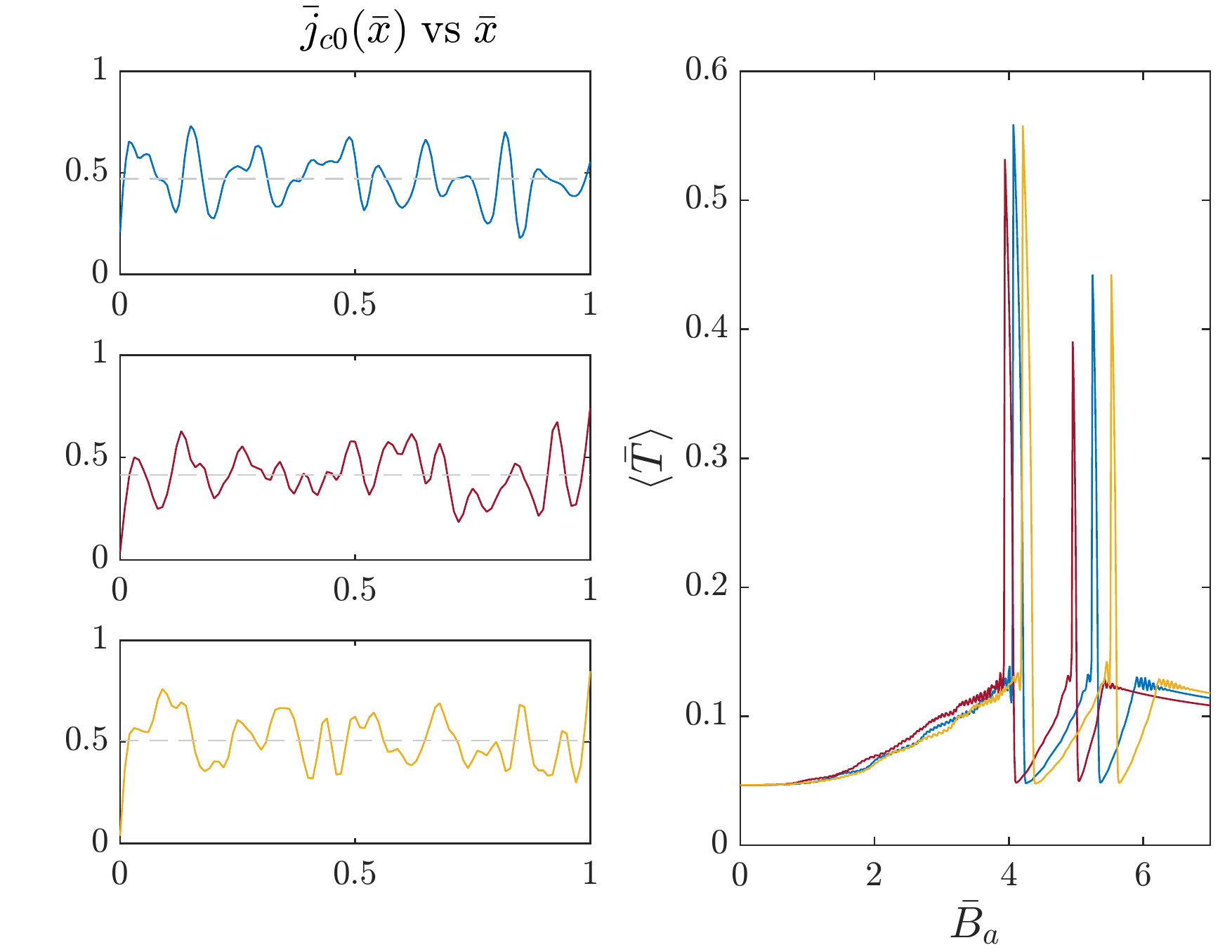}
\caption{Left. Three random spatially profiles of the current-carrying ability of a superconductor with macroscopic inhomogeneities. The dashed grey lines show the spatial average of $\bar{j}_{c0}(\bar{x})$ of each profile. Right. The corresponding average temperature as the external field was increased. \label{Rfig4}}
\end{figure}

At the left column of Fig.~\ref{Rfig4} are presented three random spatially profiles of the current-carrying ability of the superconductor with macroscopic inhomogeneities. It is exhibited only half of each profile because they are symmetric with respect to the center of the sample, the dashed grey lines show the spatial average of $\bar{j}_{c0}(\bar{x})$ of each profile. At the right of the same figure were plotted the corresponding average temperature as the external field increased, one can notice that each one presented two sharp increments independently of $\bar{j}_{c0}(\bar{x})$. Due to the randomness of $\bar{j}_{c0}(\bar{x})$ each temperature rise were slightly shifted respect each other and had different sizes, therefore, the instrinsic magnetothermal properties
of the superconductor influenced on the value of the external field where each increment occurred. We found a correlation between the average of $\bar{j}_{c0}(\bar{x})$ and the temperature rise: the smaller the average value of $\bar{j}_{c0}(\bar{x})$, the smaller the magnitude of external field where the temperature increased dramatically, this result was expected because $\bar{j}_{c0}(\bar{x})$ describes the quality of the sample with respect to its current-carrying ability. 

\section{Conclusions}
Numerical simulations were successfully performed for the study of the magnetothermal dynamics in a infinite superconducting plate with macroscopic inhomogeneities. The computing time of each simulation is quite short, of the order of seconds, despite of the stringent error tolerance, a modest computing capacity, and the interpreted data, therefore, the numerical methodology is effective for this simple geometry.

The inhomogeneities indeed affect the magnetic induction, current and temperature distributions. At the lower temperature of the thermal bath, it was notorious the effect of the inhomogeneities on the magnetic induction, however, the current density is quite more sensible to their presence. This result suggests that if it is required to detect inhomogeneities in a superconductor, it would be optimal to focus the study on the current density. 

In a thermal bath of $T_B=77$K, the spatial average of the temperature was kept uniform in a time-varying external magnetic field, independently of the inhomogeneities. On the contrary, if the superconductor is surrounded by a helium exchange gas, the temperature rise was significant. In the homogeneous sample the increment of the temperature was higher and broader than in the inhomogeneous one, however, the  presence of the inhomogeneities induced sharp increments of the temperature. 

We found that an stochastic profile for $j_{c0}$ regulates the current-carrying ability of the superconductor, which can be quantified with the average value of  $j_{c0}$, controls the magnetic field where the first magnetothermal instability occurs, and promotes the flux jumps. 

As future work, it is desirable to perform numerical simulations of more realistic geometries, for example, long superconducting bars with different cross sections. In these cases, it is required a two-dimensional model for the magnetic and thermal properties of the superconducting material. Additionally, a thorough study of thermomagnetic instabilities in inhomogeneous media is required.
\section*{Acknowledgements}
P. L. V. N. thanks CONACYT for the postdoctoral scholarship, as well as UABJO for the facilities granted. C. E. A. C. appreciates the hospitality and support provided by the UABJO.

%

%
\bibliography{references_SPI}
\end{document}